\documentclass[aps,prl,twocolumn,superscriptaddress,floatfix]{revtex4-1}

\usepackage{amsmath}
\usepackage{graphicx}
\usepackage{mathtools}
\usepackage{xcolor}
\usepackage{float}
\usepackage{bm}
\usepackage[normalem]{ulem}
\usepackage{dcolumn}

\begin{document}

\title{Hidden magnetic phases in \textit{i}-MAX compounds}

\newcommand{\BGU}{Department of Physics, Ben-Gurion University of the Negev, Beer-Sheva 84105, Israel}
\newcommand{\NRCN}{Physics Department, NRCN, P.O. Box 9001, Beer Sheva, 84190, Israel}

\newcommand{\NHMFL}{National High Magnetic Field Laboratory, Tallahassee, Florida 32310, USA}

\newcommand{\Linkoping}{Materials Design, Department of Physics, Chemistry and Biology (IFM), Linkoping University, SE-58183 Linkoping, Sweden}

\author{Dror Yahav}
\affiliation{\BGU}

\author{Ariel Maniv}
\affiliation{\NRCN}

\author{Daniel Potashnikov}
\affiliation{\NRCN}

\author{Asaf Pesach}
\affiliation{\NRCN}

\author{El'ad N. Caspi}
\affiliation{\NRCN}

\author{Arneil P. Reyes}
\affiliation{\NHMFL}

\author{Quanzheng Tao}
\affiliation{\Linkoping}

\author{Johanna Rosen}
\affiliation{\Linkoping}

\author{Eran Maniv}
\affiliation{\BGU}

\date{\today}

\begin{abstract}

We uncover a high-field magnetic phase in \textit{i}-MAX compounds exhibiting a canted antiferromagnetic (AFM) order with unprecedented properties, revealed through NMR and AC susceptibility. Intriguingly, as the atomic number of  Rare Earth increases, the transition field of this canted AFM phase grows at the expense of the lower-field AFM state. Our findings point to the complexity of the magnetic structure in \textit{i}-MAX compounds, demonstrating a non-trivial evolution of their phase diagram while increasing both the atomic number of the Rare Earth element and the external field.

\end{abstract}

\maketitle

\section{INTRODUCTION}
\textit{i}-MAX compounds are in-plane ordered nanolaminated materials, based on the well known MAX phases \cite{Barsoum2000}. In MAX phases M is an early transition metal, A is an A-group element, and X can be C, N, B, and/or P \cite{dahlqvist2023max}. They have the general chemical formula M$_{n+1}$AX$_n$, where $n$ is an integer number. These compounds have recently gained much interest, as they are parent compounds for 2D MXenes, which might be potentially interesting from several technological aspects, such as spintronics (magnetic MXenes) and energy storage, including improved batteries. \cite{naguib2021ten}. 

\textit{i}-MAX compounds have the general chemical formula (M$_{2/3}$M'$_{1/3}$)$_{2}$AX where the M, and M' elements are ordered within the M plane \cite{Dahlkvist2017},\cite{tao2017two}. This ordering is accompanied by reduction in crystal symmetry from hexagonal in MAX phases to typically monoclinic, but still maintains the general MAX phase stacking of same element layers along the $c$ axis. Most importantly, unlike in the case of MAX phase compounds, it was recently found that rare-earth (RE) elements can be introduced into the \textit{i}-MAX M' site \cite{Tao2019}.
This addition into the general MAX phase compounds opens a new set of magnetic properties that is characterized by high complexity of structures and strong dependency on the specific RE element, temperature, and magnetic field \cite{Tao2019,YangA,YangB,Sun,Chen}.

As noted in Barbier et al~\cite{barbier2022magnetic}, such compounds are expected to show complicated magnetic structure, as a result of oscillating  RKKY coupling of the 4f electrons of the different RE atoms through the conducting electron sea together with possible geometrical frustration due to the triangular RE lattice. Indeed, extensive measurements, done recently, have showed signs for such a complicated magnetic phase diagram for these compounds \cite{Tao2019,barbier2022magnetic,Tao2022,Danny}. Moreover, the complexity increases when high magnetic field is applied. \textit{i}-MAX samples with the RE elements Dy, Tb, and Ho were studied under high magnetic field by bulk magnetization, specific heat, X-ray absorption near edge structure, magnetic circular dichroism (for RE=Ho, and Dy), and neutron diffraction \cite{barbier2022magnetic,Tao2022}. Maximal applied magnetic field was in the range of 6-9\,T. Both single crystal (RE=Ho, Dy;\cite{barbier2022magnetic}), and powder (Dy, and Tb;\cite{Tao2022}) samples were studied. Magnetic Field-Temperature phase diagrams were constructed for \textit{i}-MAX compounds with RE=Dy, and Ho, presenting new high field magnetic phases in both cases. However, due to poor statistics (claimed by the authors) the exact nature of the high field magnetic phase is inconclusive. In addition, the high field magnetic structure was only studied at the lowest temperature (2K) and not its temperature dependence.

NMR and frequency dependent AC susceptibility are especially useful for determining the dynamics and magnetic phase diagram of magnetic materials at high magnetic fields. These methods have yet to be used in the study of 
\textit{i}-MAX compounds. Note that high magnetic field measurements enable better understanding of magnetic materials, especially those incorporating complex magnetic behaviour such as \textit{i}-MAX compounds, as the field can serve as a tool for dis-entangling degenerated electronic states. In addition, as a local probe measurement, NMR measures the local field distribution inside the sample, via the hyperfine interaction, hence enabling to gain better understanding on the nature of magnetism of \textit{i}-MAX samples, especially at high magnetic fields. Furthermore, this method can relatively well distinguish between the main phase contribution, and any magnetic contribution that may be a result of magnetic impurities.

In the following we show AC susceptibility and NMR measurement results of \textit{i}-MAX compounds of the form (Mo$_{2/3}$RE$_{1/3}$)$_2$AlC, where RE=Gd, Dy, Ho and Er. Note that for clarity, in the following we use the term "RE-\textit{i}" instead of the exact chemical formula. Our measurements agree with the existence of high field magnetic phase for Dy-\textit{i}, and Ho-\textit{i} but clearly show dramatic extension to high temperatures (up to 50K in Ho-\textit{i}). Moreover, this phase grows in the temperature-field phase space as the RE atomic number is increased until it dominates in Er-\textit{i}. The high field magnetic phase shows intriguing properties such as increasing transition temperature, albeit smearing, when the field is increased.

\section{Methods}

\subsection{Sample characterization}

All samples used in this work are powder samples taken from batches that were prepared and characterized previously \cite{Danny}. Unfortunately, impurities exist in all RE-\textit{i} samples reported to date by all research groups that studied powder samples. Since these impurities could contain RE and/or transition metal elements, they might be magnetic, and therefore may influence the results. This issue was not addressed in previous work well enough, and we find it vital to do so here. For that reason, a specimen from each sample was remeasured by x-ray diffraction (XRD) and carefully reanalyzed for impurities existence. 

XRD measurements were performed on a Bruker D2 diffractometer, at the Nuclear Research centre - Negev, Israel, in the 2$\theta$=7-109$^o$ range and a $\Delta$2$\theta$=0.01$^o$ step. Significant impurities are listed in Table S1 in the Supplementary information. Most of these impurities have relatively low transition temperature ($<$10K), and are not expected to significantly impact the AC susceptibility and NMR measurements, since the transition temperatures of most RE-\textit{i} compounds measured are higher than those of the impurities. Two exceptions should be noted: the ferromagnetic compound GdAl$_2$ (T$_c$=170K \cite{Williams}, $\sim0.7$\,wt\%) in the Gd-\textit{i} sample, and the superconducting one Mo$_3$Al$_2$C (T$_c$=9.2K \cite{karki2010structure}), which is found in several compounds.

\subsection{Magnetic susceptibility and NMR measurements}

To thoroughly characterize the high magnetic phase diagram of \textit{i}-MAX samples, various magnetization techniques were used, including AC susceptibility and NMR. AC susceptibility was measured using the PPMS AC susceptibility option (Quantum Design) at the Physics Department of Ben-Gurion University, Israel. The measurements were done in the temperature range of T = 3 - 300K and field range of H = 0 - 9 T. For the AC susceptibility measurements, three frequencies were used: 0.75kHz, 3.5kHz and 10kHz. Anomalies in the AC susceptibility curves appear at temperatures that were determined by the maximum of the real susceptibility $\chi'(T)$, while a few data points were determined by fitting a double Gaussian to $\chi'(T)$ where the anomaly was broad. NMR measurements were done at the National High Magnetic Field Laboratory located at Tallahassee, Florida, USA. Field-swept Al-27 NMR spectra were obtained in the field range H = 0 - 17T for temperatures between T = 4-150K, in a variable temperature cryostat. The NMR spectra were determined from summing the Fourier transforms of spin-echoes and plotted as a function of field. 

\begin{figure*}[htbp]
    \centering
    \includegraphics[width=0.9\linewidth]{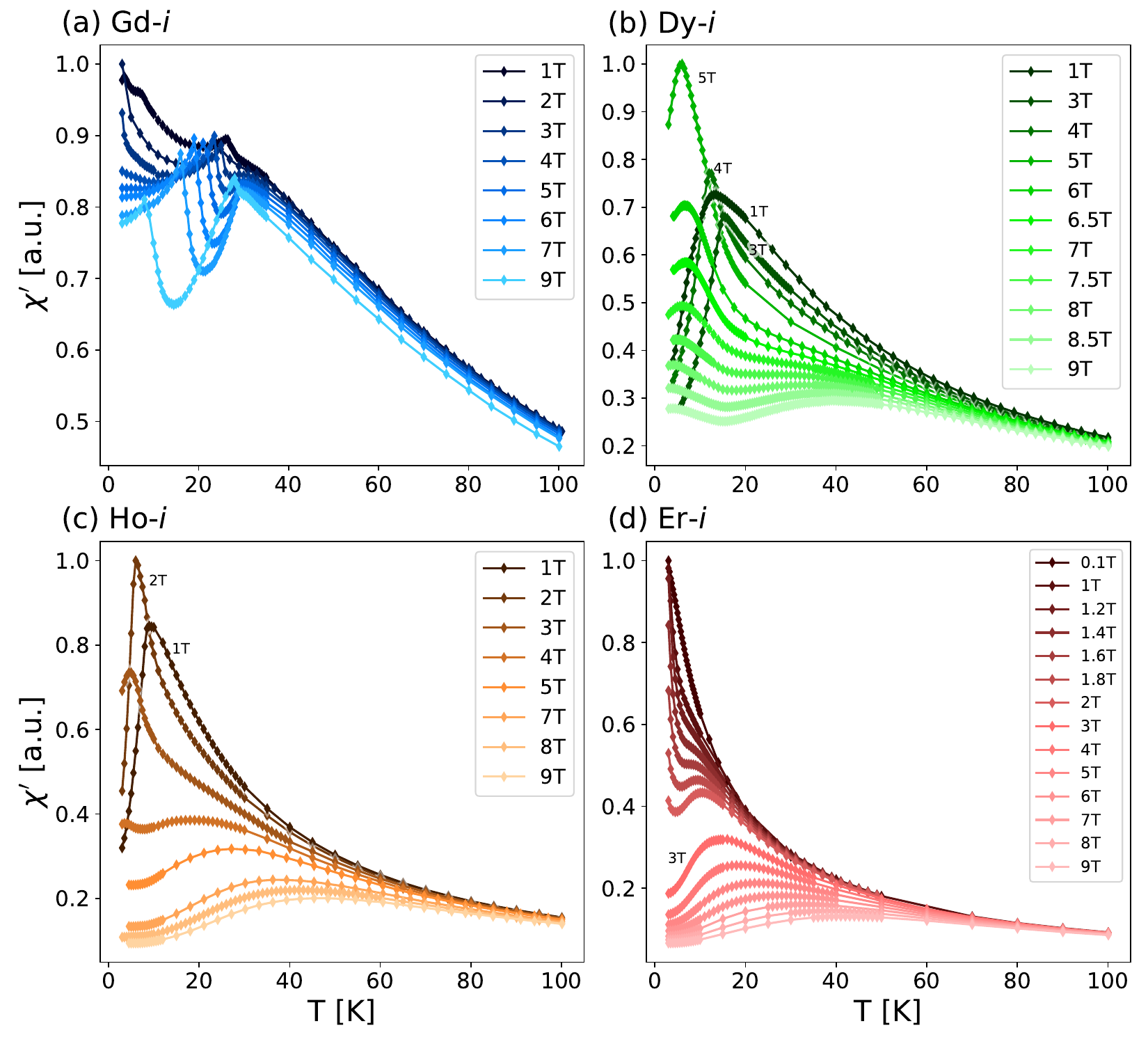}
\caption{\textbf{AC susceptibility temperature scans}. Normalized AC susceptibility measurements as a function of temperature. Upon cooling, a Curie-Weiss like behaviour is observed followed by lower temperature anomalies (a) RE=Gd (blue)  with two stable anomalies until H=9T. (b) RE=Dy (green) with a single anomaly until H=5T that is stable until H=9T. Around H=6T a second anomaly emerges that grows with the field. (c) RE=Ho with a single anomaly until H=3T, stable up to H=4T, where a second anomaly emerges that grows with the field. (d) RE=Er with a single anomaly that grows with the field. The "High field anomaly" in Dy-\textit{i}, Ho-\textit{i} and Er-\textit{i} all show a similar behaviour with a new anomaly entering that increases with the field, becomes broader while decreasing the signal. All measurements were done at a frequency of F=10kHz for various fields. Most measurements were done in the temperature range of T=3-100K. Measurements were also taken at a frequency of F=0.75kHz with a similar response.} 
    \label{fig:TS}
\end{figure*}

\section{Results}

\subsection{AC Susceptibility}

AC susceptibility results of Gd-\textit{i}, Dy-\textit{i}, Ho-\textit{i} and Er-\textit{i} samples are marked throughout the paper by blue, green, orange and red colors, respectively. Typical temperature dependence measurements are shown in Fig.~\ref{fig:TS} for several fields.  

 Upon cooling down, the AC susceptibilities in all samples increase in a Curie-Weiss like behaviour, followed by lower temperature anomalies. These anomalies can be classified according to their width in the temperature dimension. The sharp anomalies, exist also in zero field, generally refer to spontaneous magnetic ordering, or to field induced magnetic transitions. The broad anomalies may relate to crystalline electric field (CEF) effects. However, this explanation is highly unlikely, for various reasons. Firstly, since usually the peak position of AC susceptibility measurements shift to lower temperatures while increasing the magnetic field\cite{Lee}, contrary to the present results. Also, it contradicts the other measurements, especially NMR (see below) and also Neutron scattering\cite{barbier2022magnetic} , all pointing to the existence of an ordered high field magnetic phase. In addition, clearly, an evolution of the magnetic field-temperature phase diagram is observed as the RE atomic number increases (Fig.~\ref{fig:TS}). Starting from a sharp single transition ($\sim26K$ at zero field)\cite{Tao2019} for Gd-\textit{i}, upon increasing the magnetic field, it shifts to lower temperatures until it disappears completely. Near H=3T, another sharp magnetic transition appears. The transition temperature of this phase does not change significantly while increasing the field. This low temperature sharp transition can also be seen for Dy-\textit{i}, where, upon raising the magnetic field, it first shifts to lower temperature, increases to higher temperatures and then shifts back to lower temperatures. 
 
 Note also that at zero field (Fig. S1 and Fig. S2 in the Supplementary information) there is another sharp drop in the AC susceptibility of Gd-\textit{i}, and Dy-\textit{i} below $T\sim10K$. This drop resembles a superconducting transition, and hence may refer to remnants of the superconducting Mo$_3$Al$_2$C or Mo$_2$C impurities. Next, a high field broad anomaly appears for Dy-\textit{i} near H=6T. Surprisingly, increasing the field results in an increase in the temperature where the anomaly gains its full height. The higher field anomaly is also evidenced in Ho-\textit{i}, starting near H=4T, while the lower magnetic transition is truncated near this fields. 
 
 Finally, Er-\textit{i} shows a single broad magnetic anomaly. Surprisingly, the known low temperature and low field transition ($\sim5K$ at zero field, a bit higher than 3.6K reported in~\cite{Tao2019}) diminishes almost immediately while increasing the field (Fig. S4), after-which a high field anomaly (starting near H=2T) emerges, where its peak temperature increases while increasing the field. Note that such behaviour has also been evidenced for Ho-\textit{i} for a much shorter field range, as also shown below in Fig.~\ref{fig:PD}.

\begin{figure*}[t]
    \centering
    \includegraphics[width=0.9\linewidth]{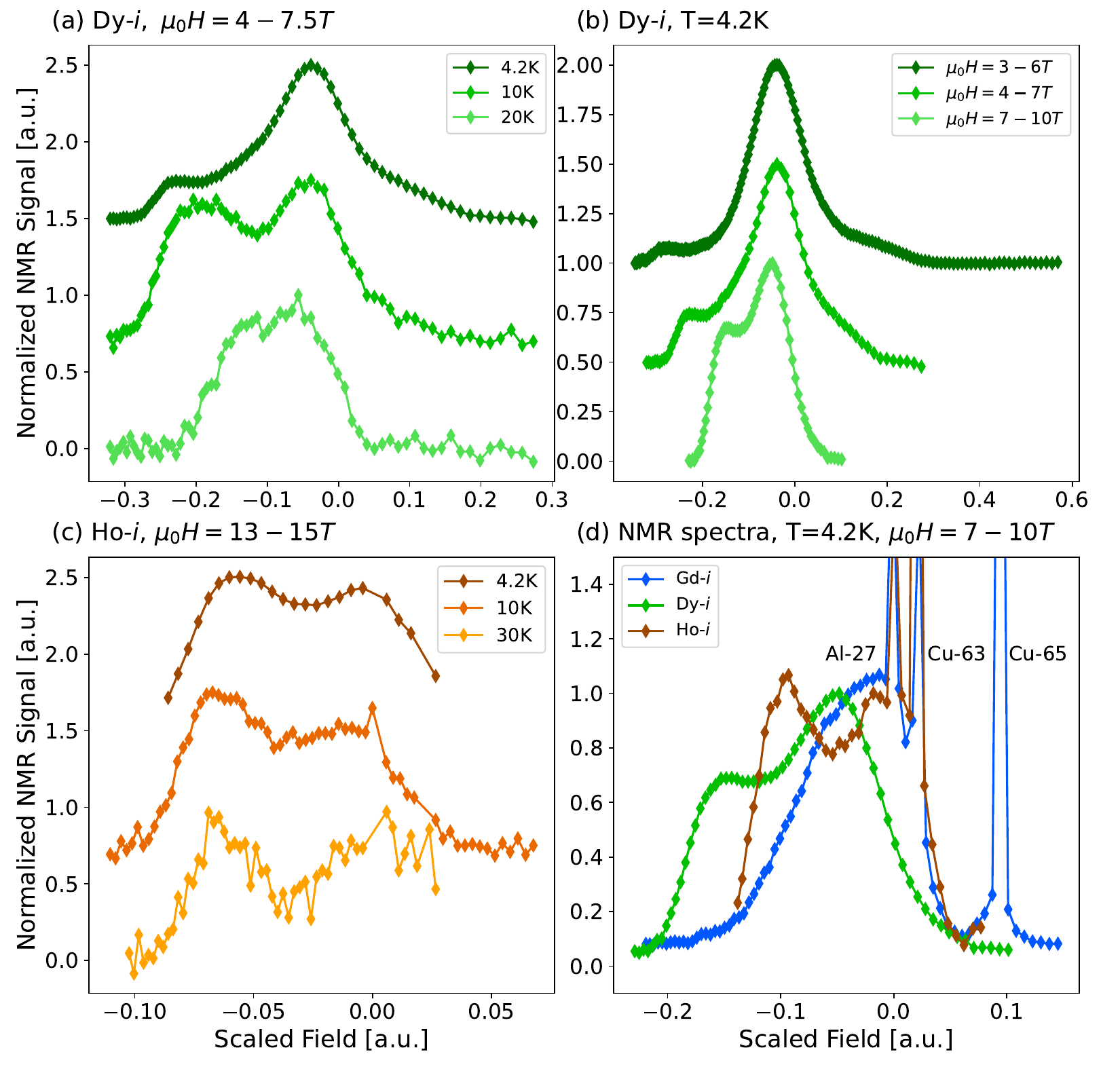}
\caption{\textbf{NMR magnetic field scans}.
(a) Field sweep Dy-\textit{i} measurements. Various Dy-\textit{i} field sweep measurements, after field scaling, taken at different temperatures. All measurements were obtained at a frequency of 56.5MHz and in the field range H=4-7.5T. The presented data is offset on the y-axis for clarity.
(b) NMR field sweep measurements of Dy-\textit{i}. All measurements were done at a temperature of T=4.2K. The presented data is offset on the y-axis for clarity.
(c) NMR measurements Ho-\textit{i}, temperature dependence. All measurements done at a frequency of F=148MHz, corresponding to a field range between H=13-15T. The presented data is offset on the y-axis for clarity.
(d) {Normalized Al-27 NMR spectra in all samples measured plotted as a function of the rescaled field $H_N$.} All measurements performed at T=4.2K for a field span between 7-10T. The sharp peaks seen in the spectra are attributed to either Copper (coming from the NMR coil) or residual Aluminum in the samples.}
    \label{fig:NMR}
\end{figure*}

\subsection{NMR}
 Fig.~\ref{fig:NMR} shows various NMR measurements done on several samples, focusing on Dy-\textit{i} (Fig.~\ref{fig:NMR}a,b), Ho-\textit{i} (Fig.~\ref{fig:NMR}c), and a comparison of NMR spectra among different samples (Fig.~\ref{fig:NMR}d). The spectra are plotted as a function of the re-scaled field:
\begin{equation}
H_N=(\omega_0/\gamma\cdot H)-1
\label{NMR_Eq}
\end{equation}
Where $\omega$$_0$ is the spectrometer frequency and $\gamma$ is the gyromagnetic ratio of Al-27 nucleus ($\gamma$=11.094MHz/T). Note that this plot automatically captures the value of the effective Knight shift relative to a bare nucleus. All measurements shown in Fig.~\ref{fig:NMR}d were made at T=4.2K, while sweeping the field in the range H=7-10T.

A clear double-peak line-shape can be recognized from Fig.~\ref{fig:NMR} only for several measurements of Dy-\textit{i} and for all measurements of Ho-\textit{i}. In contrast, Gd-\textit{i} shows a broad, probable double peak line-shape (Fig.~\ref{fig:NMR}d). Note also that the double-peak structure observed are negatively shifted. In addition, for Dy-\textit{i}, a triple peak can be seen for the lower field measurements, as evidenced by the low field (near $H_N$=0.2, H=3-6T) shoulder of the central peak (Fig.~\ref{fig:NMR}b). This triple peak, approximately symmetric line-shape, turns to an asymmetric, double peak structure while increasing the field. Finally, Ho-\textit{i} exhibits a clear, relatively symmetric, double-peak structure, negatively field shifted, at high external fields (13-15T), up to a temperature of T=30K (Fig.~\ref{fig:NMR}c). The meaning of these NMR line-shapes is explained below in the Discussion section.  

\begin{figure*}[ht]
    \centering
    \includegraphics[width=0.9\linewidth]{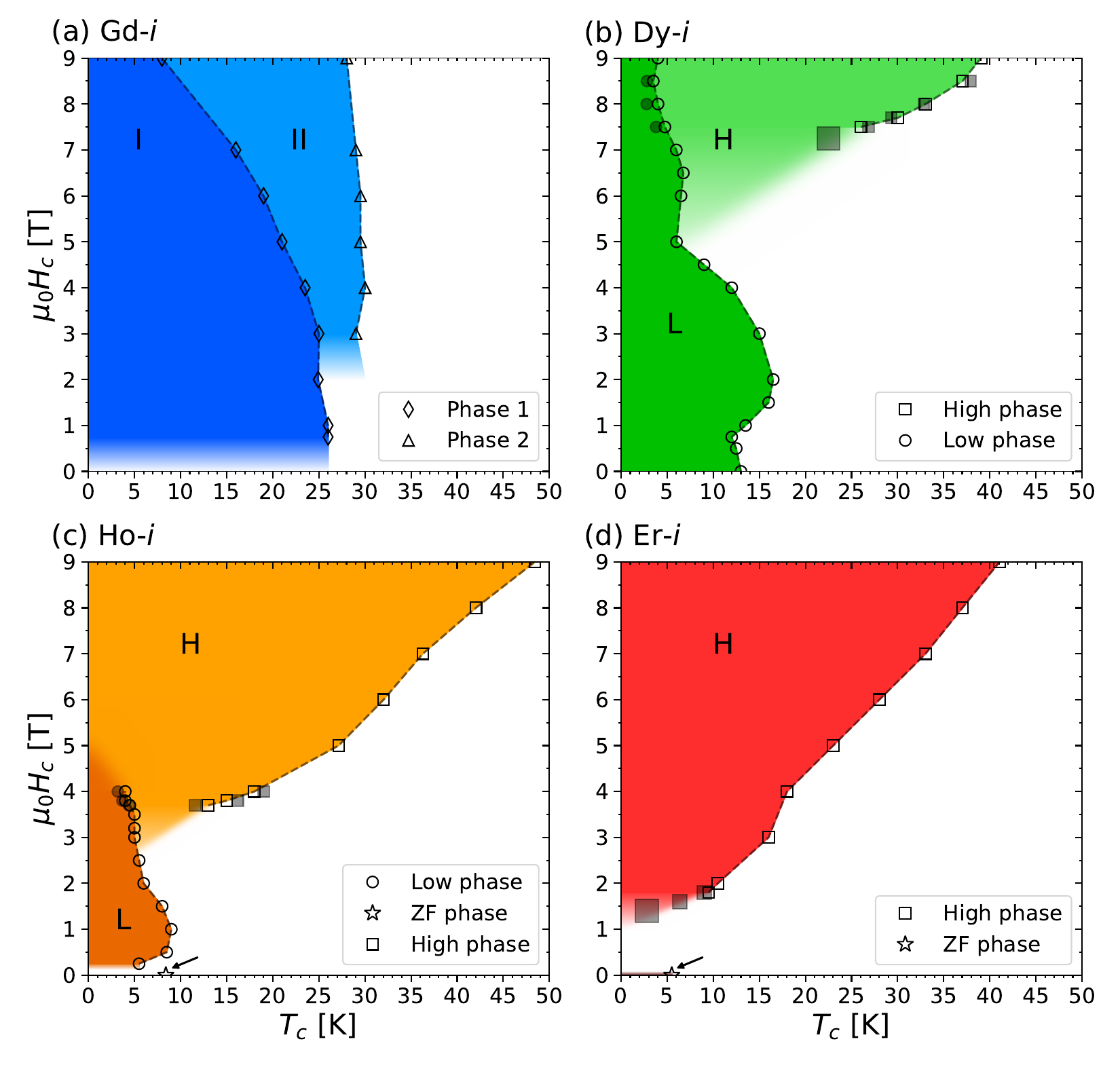}
\caption{\textbf{Phase Diagram}. Phase Diagram for each of the RE-\textit{i} samples with RE= (a) Gd, (b) Dy, (c) Ho and (d) Er marked by blue, green, orange and red colors respectively. A clear evolution of the magnetic structure of matter while increasing RE is evidenced. Specifically, from two competing magnetic phases (Gd-\textit{i}, phases 1 and 2) existing for a large portion of the H-T phase, into a formation of a new high field / high temperature (high BT, marked as H) phase is shown (Dy-\textit{i}/Ho-\textit{i}/Er-\textit{i}), where the high field phase seems to come on the expense of the lower one (marked as L), until for Er-\textit{i} the low phase totally disappears. The arrow in panels (c) and (d) refers to the ZF phase seen in Fig. S3 and Fig. S4. All diagrams were plotted for a frequency of F=10kHz. The gradient of the colors and the blur are a guide to the eye showing where the phase transition is still growing and there is a large uncertainty about the location of the phase transition. The size of the markers is larger or equal to the error-bars for simplicity.} 
    \label{fig:PD}
\end{figure*}

\section{Discussion}

The field-temperature magnetic phase diagram of the RE-\textit{i} compounds can be constructed from the AC susceptibility measurements by finding the maximum points of each measurement (Fig.~\ref{fig:PD}). Here we assume that all anomalies measured by AC susceptibility are due to magnetic phase transitions. In Fig.~\ref{fig:PD} the open data points were taken by the AC susceptibility's maximum, and the closed (transparent) data points were taken from fitting double Gaussian to the broad peak data. The difference between the phase diagrams of the different compounds measured is striking. Especially, the changes initiated by increasing the external field are dramatic, and seem to strongly depend on RE.

Gd-\textit{i} has the lightest RE element measured in this work. The zero field magnetic properties of Gd-\textit{i} are relatively well-known \cite{Tao2019}, \cite{Danny}. It orders magnetically in zero magnetic field below T=26K \cite{Tao2019}. This magnetic state was shown to be a commensurate AFM~\cite{Danny}. The transition temperature of this magnetic state is suppressed by increasing the magnetic field. At H=3T another magnetic transition appears near T=30K. This transition is stable at least up to a field of H=10T, as measured by AC susceptibility (up to 9T, Fig.~\ref{fig:TS}a), and NMR (Fig.~\ref{fig:NMR}d). Note that for an AFM structure, we expect to measure a NMR double-peak structure. However, the broad non-symmetric NMR line-shape measured is probably the combination of two broad peaks together with a significant ferromagnetic component emerging either from GdAl$_2$ magnetic impurity and from the canted AFM state. 

Interestingly, while examining the higher RE elements starting with Dy-\textit{i}, we reveal a very different picture than what is observed for Gd-\textit{i}. Initially, at low fields, the Dy-\textit{i} shows a dual transition at T=9K, 13K at zero magnetic field seen in Fig. S2. The T=13K transition matches the transition reported in the literature \cite{Tao2019}, \cite{barbier2022magnetic}, however the T=9K transition might be due to the Mo$_3$Al$_2$C parasitic impurity, which turns superconducting at T=9.2K. Note also that no match for the T=16K literature transition\cite{Tao2019} was found in the current AC susceptibility measurements, maybe due to smearing of that transition, as powder samples were measured. In addition, the transitions mentioned in the literature show slim changes in the signal of the magnetization charts\cite{Tao2019}, \cite{barbier2022magnetic}. The lower transition is observed only up to a field of H=0.5T. The higher field anomaly stretches at least up to H=9T. Note that at a field of H=5T a significant drop in transition temperature has been observed for the low field transition, probably due to the magnetic moment saturation reported in \cite{barbier2022magnetic}. We claim that this presumable moment saturation is actually a transition to another magnetic state, as will be explained below. Specifically and surprisingly, at the same point (near H=5T) another high field magnetic anomaly emerges. The transition temperature of this purported new magnetic state is enhanced at higher fields. This behaviour is opposite to both Gd-\textit{i} and Tb-\textit{i} (see \cite{Preprint_Tb}, results not shown in this manuscript), which show a decrease in the anomaly temperature at high fields. Also, NMR measurements (Fig. \ref{fig:NMR}b) show a dramatic change of line-shape between the two phases, from a nearly symmetric triple-peak line-shape for low fields (low transition temperature phase), to a negatively shifted re-scaled field, double peak structure for the high field phase, implying a canted non-symmetric AFM. Specifically, as seen in Fig. \ref{fig:NMR}b, the large peak near the resonance field for H=3-6T field sweep measurements is probably due to the spin portion of the lattice, which is already aligned in the external field direction (canting AFM), while the two side peaks originate from the AFM state. While increasing the field, a double-peak emerges, indicating reduced AFM hyperfine component (Fig. \ref{fig:NMR}b). This is accompanied by a slight increase in the FM (more negative shift) component which suggests further canting of the spins as the temperature is increased, and the sample shifts from the lower magnetic phase to the high field phase. A similar change from low phase (AFM, T=4.2K) to high phase (canted AFM, T=10K) can also be recognized in Fig. \ref{fig:NMR}a, where increasing the temperature changes the Dy-\textit{i} magnetic phase (see also Fig. \ref{fig:PD}b).

The Ho-\textit{i} data shows a transition temperature of T=8.5K at zero field, identical to the transition temperature reported in \cite{YangA}, and close to that reported in \cite{Tao2019}, and \cite{barbier2022magnetic}. This transition terminates at H=0.01T and then re-enters at H=0.3T, existing up to a field of H $\approx$ 4T, near which a high field magnetic transition emerges at T=13K. The transition temperature of this magnetic phase then increases with field, similar to the Dy-\textit{i} high field magnetic transition. Note that NMR measurements show the existence of this magnetic state at least up to a field of H=15T at T=4.2K (See Fig. \ref{fig:NMR}c).

Finally, Er-\textit{i} exhibits a low temperature transition (T=5.5K) at zero field, similar to the T=3.6K transition noted in the literature~\cite{Tao2019}. However, this transition disappears while increasing the field to an external field of merely H=0.04T. At a field of H=2T the high field phase emerges, where increasing the field up to (at-least) H=9T implies an increase in the transition temperature, similar to Dy-\textit{i} and Ho-\textit{i}.

\begin{figure}[htbp]
    \centering
    \includegraphics[width=\linewidth]{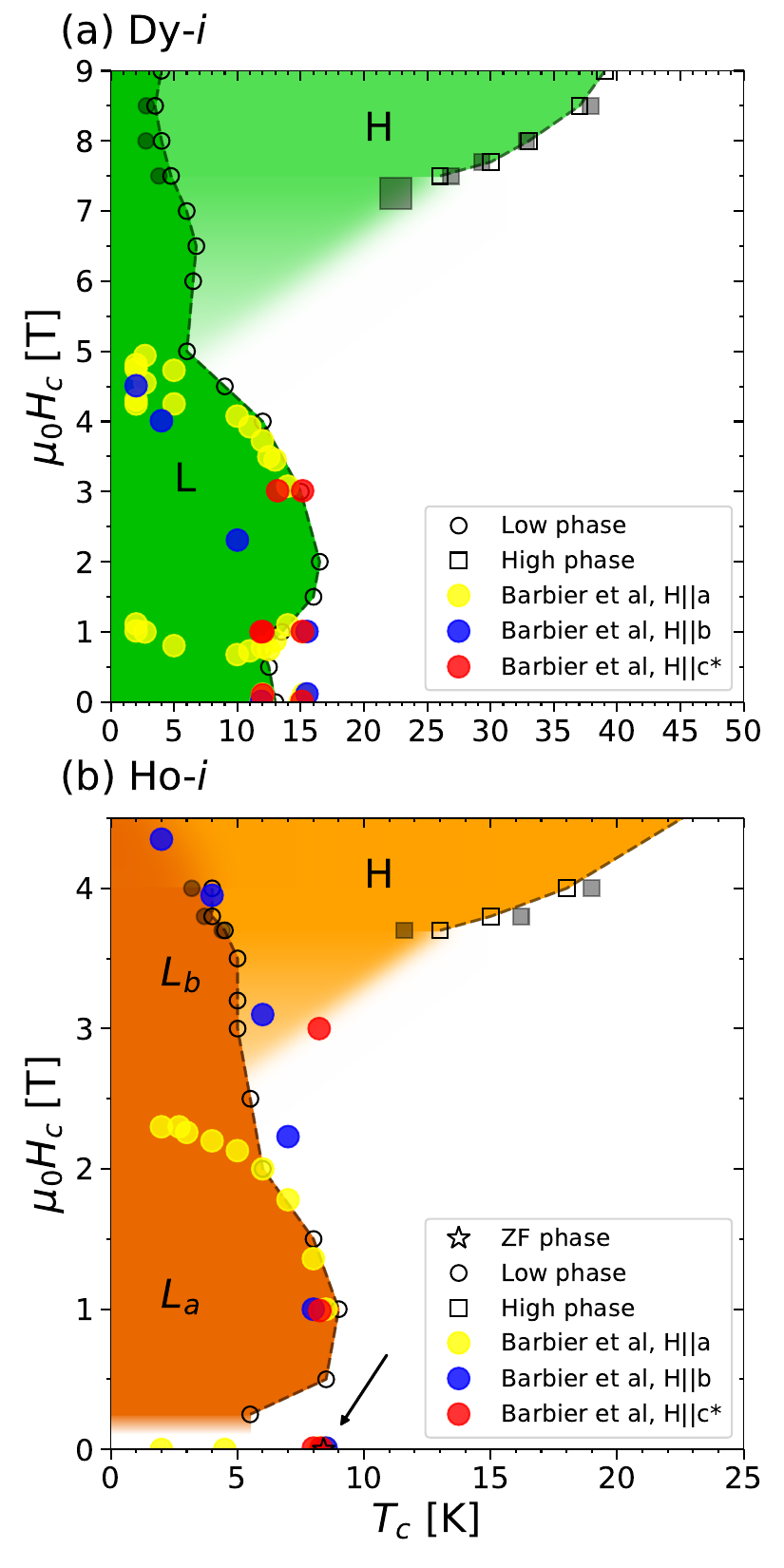}
\caption{\textbf{Dy-\textit{i} and Ho-\textit{i} phase diagram comparison}. AC susceptibility data measured for powder (a) Dy-\textit{i} and (b) Ho-\textit{i} samples showing the zero magnetic field (ZF), Low and High temperature/magnetic field phases (open star, circles and squares respectively). Thermodynamic and diffraction data
measured for single crystal Ho-\textit{i} samples in a, b and c axes (yellow, blue and red circles respectively), taken from Barbier et al~\cite{barbier2022magnetic}. The comparison of the present results to the previously published data \cite{barbier2022magnetic} is remarkable, showing the full axis dependent phase diagram in the powder measurement and the formation of the High temperature/magnetic field phase. In the Ho-\textit{i} sample there is a clear distinction between the low phase in a or b direction, hence $L_a$ and $L_b$ referring to different contributes from a, b axes to the powder diagram. The Dy-\textit{i} samples are more complicated hence was left with a single "L" marker.} 
    \label{fig:Fitted}
\end{figure}

All the above findings can be summarized in the high field phase diagrams of Dy-\textit{i} and Ho-\textit{i} (Fig.~\ref{fig:Fitted}). A comparison with previous literature results, taken using single crystals, is also presented, showing, in general, good agreement for the low field phase  \cite{barbier2022magnetic}. Deviation from the literature\cite{barbier2022magnetic} is observed mainly for the case where previous measurements were done with magnetic field aligned along the a-axis, which serves as the easy axis in these compounds. This transition is probably smeared out when powder samples are measured, as is the case in this work. Note, that the high field, high temperature phase is observed here for the first time for the RE-\textit{i}-MAX phases. 

These findings point to the emergence of a novel high field phase. The universality of the high magnetic phase for various \textit{i}-MAX compounds is clear. This new magnetic phase has peculiar features, very different from the low magnetic phases already noted in previous studies \cite{Tao2019}, \cite{barbier2022magnetic}. Specifically, the high phase, which has a significant FM component is different from the low magnetic phases. While the low magnetic phases show a typical behaviour, in which the transition temperature drops while increasing the magnetic field, the high phase shows the opposite. In addition, the high phase transition is smeared while increasing the field. Note that the emergence of this high field phase might be due to a competition between different magnetic phases, similar to the ones shown in~\cite{EM2}. There, while increasing the field, one magnetic order changes to another, more energetically favorable. Another hint on the nature of this high field phase can be found in the evolution of the Ho-\textit{i} phase diagram as a function of field measured by neutron diffraction \cite{barbier2022magnetic}. Starting with a AFM structure, a FM component emerges for relatively low magnetic field (1T). This component develops with increased field until along the a axis a full FM phase is suggested for H$>$2T. Therefore, we hypothesize that the high phase found here has a large FM component, that becomes more prominent as the field increases. This magnetic transition appears at higher temperatures as the strength of the applied field increases probably due to stabilization effect of the FM component by field application. The NMR measurements support the above findings, as a significant double-peak structure, attributed to magnetically ordered canted AFM state (see also ~\cite{EM}), has been shown up to a field of H=17T for Dy-\textit{i} and Ho-\textit{i}.

\section{Conclusions}

Our investigation of several \textit{i}-MAX compounds using a variety of experimental techniques has revealed intriguing insights into their magnetic behavior. We uncovered a complex magnetic phase diagram featuring the re-entrance of magnetic phases and the existence of a high-field phase. Notably, we observed contrasting behavior between Gd-\textit{i} and Tb-\textit{i}~\cite{Preprint_Tb} compounds, where the ordered magnetic state weakens with increasing field, compared to Dy-\textit{i}, Ho-\textit{i}, and Er-\textit{i}, where the transition temperature increases at least up to fields of 9T. This increase is likely linked to the contraction of the unit cell, which suppresses the low-field phase and enhances the high-field phase. It is followed by the introduction of canted AFM order, with probable increase in the FM component significance, which survives even at fields as high as 17T (for Ho-\textit{i}, see Fig \ref{fig:NMR}c). This points to the uniqueness of the high-field phase in these materials.
Our results indicate that a competition exists between different magnetic phases, with one phase becoming energetically more favorable as the external field increases, similar to the one shown in~\cite{EM2}. This behavior may explain the observed disappearance and subsequent re-emergence of magnetic phases with increasing field.

Based on our findings and earlier studies~\cite{Tao2019, barbier2022magnetic,Tao2022}, we propose the following picture of the magnetic state in 
(Mo$_{2/3}$RE$_{1/3}$)$_2$AlC compounds. In the ground state, the magnetic configuration is complex due to the geometrical frustration of RE atoms, leading to an antiferromagnetic spin density wave (AFM SDW) for most compounds at zero field~\cite{Danny}. As the field increases, a canted AFM lattice emerges, with a stabilized FM component that aligns along the a easy axis and persists even at high fields, up to 17T as evidenced in Ho-\textit{i}.
Furthermore, increasing the atomic number of the RE elements leads to a decrease in inter-atomic distance~\cite{Tao2019}. From the results presented here, as well as from the zero-field transition temperatures presented elsewhere \cite{Tao2019,Danny}, it is clear that the interaction strength of the low field phase diminishes as well, starting from Dy-\textit{i}. These findings needs to be further explored, theoretically as well as experimentally, since no direct exchange mechanism is expected. This means that the strength in magnetic interaction do not have to decrease with decreasing of lattice dimensions, and a more complicated correlation is expected.
The application of an increasing magnetic field also leads to a reduction in the strength of the low-field phases, which eventually vanish for Er-\textit{i}, while the high-field phase becomes increasingly stabilized (for Dy-\textit{i} and above). This behavior likely stems from the stabilization of the FM component along the a easy axis under high magnetic fields.
Additional measurements, particularly high-field neutron diffraction, are needed to precisely characterize the high-field behavior of \textit{i}-MAX compounds and to further clarify the nature of these intriguing magnetic phases.

\

\textbf{Acknowledgements}
Work by D.Y. and E.M. was performed with support from the European Research Council grant no. ERC-101117478, the Israeli Science Foundation grant no. ISF-885/23 and the PAZY foundation grant no. 412/23.
JR acknowledges funding from the Knut and Alice Wallenberg (KAW) Foundation for a Scholar Grant (2019.0433).


\newpage

\begin{thebibliography}{19}%
\makeatletter
\providecommand \@ifxundefined [1]{%
 \@ifx{#1\undefined}
}%
\providecommand \@ifnum [1]{%
 \ifnum #1\expandafter \@firstoftwo
 \else \expandafter \@secondoftwo
 \fi
}%
\providecommand \@ifx [1]{%
 \ifx #1\expandafter \@firstoftwo
 \else \expandafter \@secondoftwo
 \fi
}%
\providecommand \natexlab [1]{#1}%
\providecommand \enquote  [1]{``#1''}%
\providecommand \bibnamefont  [1]{#1}%
\providecommand \bibfnamefont [1]{#1}%
\providecommand \citenamefont [1]{#1}%
\providecommand \href@noop [0]{\@secondoftwo}%
\providecommand \href [0]{\begingroup \@sanitize@url \@href}%
\providecommand \@href[1]{\@@startlink{#1}\@@href}%
\providecommand \@@href[1]{\endgroup#1\@@endlink}%
\providecommand \@sanitize@url [0]{\catcode `\\12\catcode `\$12\catcode `\&12\catcode `\#12\catcode `\^12\catcode `\_12\catcode `\%12\relax}%
\providecommand \@@startlink[1]{}%
\providecommand \@@endlink[0]{}%
\providecommand \url  [0]{\begingroup\@sanitize@url \@url }%
\providecommand \@url [1]{\endgroup\@href {#1}{\urlprefix }}%
\providecommand \urlprefix  [0]{URL }%
\providecommand \Eprint [0]{\href }%
\providecommand \doibase [0]{http://dx.doi.org/}%
\providecommand \selectlanguage [0]{\@gobble}%
\providecommand \bibinfo  [0]{\@secondoftwo}%
\providecommand \bibfield  [0]{\@secondoftwo}%
\providecommand \translation [1]{[#1]}%
\providecommand \BibitemOpen [0]{}%
\providecommand \bibitemStop [0]{}%
\providecommand \bibitemNoStop [0]{.\EOS\space}%
\providecommand \EOS [0]{\spacefactor3000\relax}%
\providecommand \BibitemShut  [1]{\csname bibitem#1\endcsname}%
\let\auto@bib@innerbib\@empty
\bibitem [{\citenamefont {Barsoum}(2000)}]{Barsoum2000}%
  \BibitemOpen
  \bibfield  {author} {\bibinfo {author} {\bibfnamefont {M.~W.}\ \bibnamefont {Barsoum}},\ }\href@noop {} {\bibfield  {journal} {\bibinfo  {journal} {Progress in Solid State Chemistry}\ }\textbf {\bibinfo {volume} {28}},\ \bibinfo {pages} {201} (\bibinfo {year} {2000})}\BibitemShut {NoStop}%
\bibitem [{\citenamefont {Dahlqvist}\ \emph {et~al.}(2023)\citenamefont {Dahlqvist}, \citenamefont {Barsoum},\ and\ \citenamefont {Rosen}}]{dahlqvist2023max}%
  \BibitemOpen
  \bibfield  {author} {\bibinfo {author} {\bibfnamefont {M.}~\bibnamefont {Dahlqvist}}, \bibinfo {author} {\bibfnamefont {M.~W.}\ \bibnamefont {Barsoum}}, \ and\ \bibinfo {author} {\bibfnamefont {J.}~\bibnamefont {Rosen}},\ }\href@noop {} {\bibfield  {journal} {\bibinfo  {journal} {Materials Today}\ } (\bibinfo {year} {2023})}\BibitemShut {NoStop}%
\bibitem [{\citenamefont {Naguib}\ \emph {et~al.}(2021)\citenamefont {Naguib}, \citenamefont {Barsoum},\ and\ \citenamefont {Gogotsi}}]{naguib2021ten}%
  \BibitemOpen
  \bibfield  {author} {\bibinfo {author} {\bibfnamefont {M.}~\bibnamefont {Naguib}}, \bibinfo {author} {\bibfnamefont {M.~W.}\ \bibnamefont {Barsoum}}, \ and\ \bibinfo {author} {\bibfnamefont {Y.}~\bibnamefont {Gogotsi}},\ }\href@noop {} {\bibfield  {journal} {\bibinfo  {journal} {Advanced Materials}\ }\textbf {\bibinfo {volume} {33}},\ \bibinfo {pages} {2103393} (\bibinfo {year} {2021})}\BibitemShut {NoStop}%
\bibitem [{\citenamefont {Dahlqvist}\ \emph {et~al.}(2017)\citenamefont {Dahlqvist}, \citenamefont {Lu}, \citenamefont {Meshkian}, \citenamefont {Tao}, \citenamefont {Hultman},\ and\ \citenamefont {Ros{\'e}n}}]{Dahlkvist2017}%
  \BibitemOpen
  \bibfield  {author} {\bibinfo {author} {\bibfnamefont {M.}~\bibnamefont {Dahlqvist}}, \bibinfo {author} {\bibfnamefont {J.}~\bibnamefont {Lu}}, \bibinfo {author} {\bibfnamefont {R.}~\bibnamefont {Meshkian}}, \bibinfo {author} {\bibfnamefont {Q.}~\bibnamefont {Tao}}, \bibinfo {author} {\bibfnamefont {L.}~\bibnamefont {Hultman}}, \ and\ \bibinfo {author} {\bibfnamefont {J.}~\bibnamefont {Ros{\'e}n}},\ }\href@noop {} {\bibfield  {journal} {\bibinfo  {journal} {Science Advances}\ }\textbf {\bibinfo {volume} {3}},\ \bibinfo {pages} {e1700642} (\bibinfo {year} {2017})}\BibitemShut {NoStop}%
\bibitem [{\citenamefont {Tao}\ \emph {et~al.}(2017)\citenamefont {Tao}, \citenamefont {Dahlqvist}, \citenamefont {Lu}, \citenamefont {Kota}, \citenamefont {Meshkian}, \citenamefont {Halim}, \citenamefont {Palisaitis}, \citenamefont {Hultman}, \citenamefont {Barsoum}, \citenamefont {Persson} \emph {et~al.}}]{tao2017two}%
  \BibitemOpen
  \bibfield  {author} {\bibinfo {author} {\bibfnamefont {Q.}~\bibnamefont {Tao}}, \bibinfo {author} {\bibfnamefont {M.}~\bibnamefont {Dahlqvist}}, \bibinfo {author} {\bibfnamefont {J.}~\bibnamefont {Lu}}, \bibinfo {author} {\bibfnamefont {S.}~\bibnamefont {Kota}}, \bibinfo {author} {\bibfnamefont {R.}~\bibnamefont {Meshkian}}, \bibinfo {author} {\bibfnamefont {J.}~\bibnamefont {Halim}}, \bibinfo {author} {\bibfnamefont {J.}~\bibnamefont {Palisaitis}}, \bibinfo {author} {\bibfnamefont {L.}~\bibnamefont {Hultman}}, \bibinfo {author} {\bibfnamefont {M.~W.}\ \bibnamefont {Barsoum}}, \bibinfo {author} {\bibfnamefont {P.~O.}\ \bibnamefont {Persson}},  \emph {et~al.},\ }\href@noop {} {\bibfield  {journal} {\bibinfo  {journal} {Nature communications}\ }\textbf {\bibinfo {volume} {8}},\ \bibinfo {pages} {14949} (\bibinfo {year} {2017})}\BibitemShut {NoStop}%
\bibitem [{\citenamefont {Tao}\ \emph {et~al.}(2019)\citenamefont {Tao}, \citenamefont {Lu}, \citenamefont {Dahlqvist}, \citenamefont {Mockute}, \citenamefont {Calder}, \citenamefont {Petruhins}, \citenamefont {Meshkian}, \citenamefont {Rivin}, \citenamefont {Potashnikov}, \citenamefont {Caspi} \emph {et~al.}}]{Tao2019}%
  \BibitemOpen
  \bibfield  {author} {\bibinfo {author} {\bibfnamefont {Q.}~\bibnamefont {Tao}}, \bibinfo {author} {\bibfnamefont {J.}~\bibnamefont {Lu}}, \bibinfo {author} {\bibfnamefont {M.}~\bibnamefont {Dahlqvist}}, \bibinfo {author} {\bibfnamefont {A.}~\bibnamefont {Mockute}}, \bibinfo {author} {\bibfnamefont {S.}~\bibnamefont {Calder}}, \bibinfo {author} {\bibfnamefont {A.}~\bibnamefont {Petruhins}}, \bibinfo {author} {\bibfnamefont {R.}~\bibnamefont {Meshkian}}, \bibinfo {author} {\bibfnamefont {O.}~\bibnamefont {Rivin}}, \bibinfo {author} {\bibfnamefont {D.}~\bibnamefont {Potashnikov}}, \bibinfo {author} {\bibfnamefont {E.~N.}\ \bibnamefont {Caspi}},  \emph {et~al.},\ }\href@noop {} {\bibfield  {journal} {\bibinfo  {journal} {Chemistry of Materials}\ }\textbf {\bibinfo {volume} {31}},\ \bibinfo {pages} {2476} (\bibinfo {year} {2019})}\BibitemShut {NoStop}%
\bibitem [{\citenamefont {Yang}(2021{\natexlab{a}})}]{YangA}%
  \BibitemOpen
  \bibfield  {author} {\bibinfo {author} {\bibfnamefont {e.~a.}\ \bibnamefont {Yang}, \bibfnamefont {J.}},\ }\href@noop {} {\bibfield  {journal} {\bibinfo  {journal} {Carbon}\ }\textbf {\bibinfo {volume} {179}},\ \bibinfo {pages} {104} (\bibinfo {year} {2021}{\natexlab{a}})}\BibitemShut {NoStop}%
\bibitem [{\citenamefont {Yang}(2021{\natexlab{b}})}]{YangB}%
  \BibitemOpen
  \bibfield  {author} {\bibinfo {author} {\bibfnamefont {e.~a.}\ \bibnamefont {Yang}, \bibfnamefont {J.}},\ }\href@noop {} {\bibfield  {journal} {\bibinfo  {journal} {Carbon}\ }\textbf {\bibinfo {volume} {183}},\ \bibinfo {pages} {76} (\bibinfo {year} {2021}{\natexlab{b}})}\BibitemShut {NoStop}%
\bibitem [{\citenamefont {Sun}(2023)}]{Sun}%
  \BibitemOpen
  \bibfield  {author} {\bibinfo {author} {\bibfnamefont {S.~e.~a.}\ \bibnamefont {Sun}},\ }\href@noop {} {\bibfield  {journal} {\bibinfo  {journal} {Acta Materialia}\ }\textbf {\bibinfo {volume} {242}},\ \bibinfo {pages} {118479} (\bibinfo {year} {2023})}\BibitemShut {NoStop}%
\bibitem [{\citenamefont {Chen}(2022)}]{Chen}%
  \BibitemOpen
  \bibfield  {author} {\bibinfo {author} {\bibfnamefont {Z.~e.~a.}\ \bibnamefont {Chen}},\ }\href@noop {} {\bibfield  {journal} {\bibinfo  {journal} {Scripta Materialia}\ }\textbf {\bibinfo {volume} {213}},\ \bibinfo {pages} {114596} (\bibinfo {year} {2022})}\BibitemShut {NoStop}%
\bibitem [{\citenamefont {Barbier}\ \emph {et~al.}(2022)\citenamefont {Barbier}, \citenamefont {Wilhelm}, \citenamefont {Colin}, \citenamefont {Opagiste}, \citenamefont {Lhotel}, \citenamefont {Pinek}, \citenamefont {Kim}, \citenamefont {Braithwaite}, \citenamefont {Ressouche}, \citenamefont {Ohresser} \emph {et~al.}}]{barbier2022magnetic}%
  \BibitemOpen
  \bibfield  {author} {\bibinfo {author} {\bibfnamefont {M.}~\bibnamefont {Barbier}}, \bibinfo {author} {\bibfnamefont {F.}~\bibnamefont {Wilhelm}}, \bibinfo {author} {\bibfnamefont {C.~V.}\ \bibnamefont {Colin}}, \bibinfo {author} {\bibfnamefont {C.}~\bibnamefont {Opagiste}}, \bibinfo {author} {\bibfnamefont {E.}~\bibnamefont {Lhotel}}, \bibinfo {author} {\bibfnamefont {D.}~\bibnamefont {Pinek}}, \bibinfo {author} {\bibfnamefont {Y.}~\bibnamefont {Kim}}, \bibinfo {author} {\bibfnamefont {D.}~\bibnamefont {Braithwaite}}, \bibinfo {author} {\bibfnamefont {E.}~\bibnamefont {Ressouche}}, \bibinfo {author} {\bibfnamefont {P.}~\bibnamefont {Ohresser}},  \emph {et~al.},\ }\href@noop {} {\bibfield  {journal} {\bibinfo  {journal} {Physical Review B}\ }\textbf {\bibinfo {volume} {105}},\ \bibinfo {pages} {174421} (\bibinfo {year} {2022})}\BibitemShut {NoStop}%
\bibitem [{\citenamefont {Tao}\ \emph {et~al.}(2022)\citenamefont {Tao}, \citenamefont {Barbier}, \citenamefont {Mockute}, \citenamefont {Ritter}, \citenamefont {Salikhov}, \citenamefont {Wiedwald}, \citenamefont {Calder}, \citenamefont {Opagiste}, \citenamefont {Galera}, \citenamefont {Farle} \emph {et~al.}}]{Tao2022}%
  \BibitemOpen
  \bibfield  {author} {\bibinfo {author} {\bibfnamefont {Q.}~\bibnamefont {Tao}}, \bibinfo {author} {\bibfnamefont {M.}~\bibnamefont {Barbier}}, \bibinfo {author} {\bibfnamefont {A.}~\bibnamefont {Mockute}}, \bibinfo {author} {\bibfnamefont {C.}~\bibnamefont {Ritter}}, \bibinfo {author} {\bibfnamefont {R.}~\bibnamefont {Salikhov}}, \bibinfo {author} {\bibfnamefont {U.}~\bibnamefont {Wiedwald}}, \bibinfo {author} {\bibfnamefont {S.}~\bibnamefont {Calder}}, \bibinfo {author} {\bibfnamefont {C.}~\bibnamefont {Opagiste}}, \bibinfo {author} {\bibfnamefont {R.-M.}\ \bibnamefont {Galera}}, \bibinfo {author} {\bibfnamefont {M.}~\bibnamefont {Farle}},  \emph {et~al.},\ }\href@noop {} {\bibfield  {journal} {\bibinfo  {journal} {Journal of Physics: Condensed Matter}\ }\textbf {\bibinfo {volume} {34}},\ \bibinfo {pages} {215801} (\bibinfo {year} {2022})}\BibitemShut {NoStop}%
\bibitem [{\citenamefont {Potashnikov}\ \emph {et~al.}(2021)\citenamefont {Potashnikov}, \citenamefont {Caspi}, \citenamefont {Pesach}, \citenamefont {Tao}, \citenamefont {Ros{\'e}n}, \citenamefont {Sheptyakov}, \citenamefont {Evans}, \citenamefont {Ritter}, \citenamefont {Salman}, \citenamefont {Bonfa} \emph {et~al.}}]{Danny}%
  \BibitemOpen
  \bibfield  {author} {\bibinfo {author} {\bibfnamefont {D.}~\bibnamefont {Potashnikov}}, \bibinfo {author} {\bibfnamefont {E.}~\bibnamefont {Caspi}}, \bibinfo {author} {\bibfnamefont {A.}~\bibnamefont {Pesach}}, \bibinfo {author} {\bibfnamefont {Q.}~\bibnamefont {Tao}}, \bibinfo {author} {\bibfnamefont {J.}~\bibnamefont {Ros{\'e}n}}, \bibinfo {author} {\bibfnamefont {D.}~\bibnamefont {Sheptyakov}}, \bibinfo {author} {\bibfnamefont {H.}~\bibnamefont {Evans}}, \bibinfo {author} {\bibfnamefont {C.}~\bibnamefont {Ritter}}, \bibinfo {author} {\bibfnamefont {Z.}~\bibnamefont {Salman}}, \bibinfo {author} {\bibfnamefont {P.}~\bibnamefont {Bonfa}},  \emph {et~al.},\ }\href@noop {} {\bibfield  {journal} {\bibinfo  {journal} {Physical Review B}\ }\textbf {\bibinfo {volume} {104}},\ \bibinfo {pages} {174440} (\bibinfo {year} {2021})}\BibitemShut {NoStop}%
\bibitem [{\citenamefont {Williams}\ \emph {et~al.}(2003)\citenamefont {Williams}, \citenamefont {Shand}, \citenamefont {Pekarek}, \citenamefont {Skomski}, \citenamefont {Petkov},\ and\ \citenamefont {Leslie-Pelecky}}]{Williams}%
  \BibitemOpen
  \bibfield  {author} {\bibinfo {author} {\bibfnamefont {D.~S.}\ \bibnamefont {Williams}}, \bibinfo {author} {\bibfnamefont {P.~M.}\ \bibnamefont {Shand}}, \bibinfo {author} {\bibfnamefont {T.~M.}\ \bibnamefont {Pekarek}}, \bibinfo {author} {\bibfnamefont {R.}~\bibnamefont {Skomski}}, \bibinfo {author} {\bibfnamefont {V.}~\bibnamefont {Petkov}}, \ and\ \bibinfo {author} {\bibfnamefont {D.~L.}\ \bibnamefont {Leslie-Pelecky}},\ }\href@noop {} {\bibfield  {journal} {\bibinfo  {journal} {Phys. Rev. B}\ }\textbf {\bibinfo {volume} {68}},\ \bibinfo {pages} {214404} (\bibinfo {year} {2003})}\BibitemShut {NoStop}%
\bibitem [{\citenamefont {Karki}\ \emph {et~al.}(2010)\citenamefont {Karki}, \citenamefont {Xiong}, \citenamefont {Vekhter}, \citenamefont {Browne}, \citenamefont {Adams}, \citenamefont {Young}, \citenamefont {Thomas}, \citenamefont {Chan}, \citenamefont {Kim},\ and\ \citenamefont {Prozorov}}]{karki2010structure}%
  \BibitemOpen
  \bibfield  {author} {\bibinfo {author} {\bibfnamefont {A.}~\bibnamefont {Karki}}, \bibinfo {author} {\bibfnamefont {Y.}~\bibnamefont {Xiong}}, \bibinfo {author} {\bibfnamefont {I.}~\bibnamefont {Vekhter}}, \bibinfo {author} {\bibfnamefont {D.}~\bibnamefont {Browne}}, \bibinfo {author} {\bibfnamefont {P.}~\bibnamefont {Adams}}, \bibinfo {author} {\bibfnamefont {D.}~\bibnamefont {Young}}, \bibinfo {author} {\bibfnamefont {K.}~\bibnamefont {Thomas}}, \bibinfo {author} {\bibfnamefont {J.~Y.}\ \bibnamefont {Chan}}, \bibinfo {author} {\bibfnamefont {H.}~\bibnamefont {Kim}}, \ and\ \bibinfo {author} {\bibfnamefont {R.}~\bibnamefont {Prozorov}},\ }\href@noop {} {\bibfield  {journal} {\bibinfo  {journal} {Physical Review B}\ }\textbf {\bibinfo {volume} {82}},\ \bibinfo {pages} {064512} (\bibinfo {year} {2010})}\BibitemShut {NoStop}%
\bibitem [{\citenamefont {Lee}\ \emph {et~al.}(1999)\citenamefont {Lee}, \citenamefont {Kim}, \citenamefont {Kim}, \citenamefont {Ri}, \citenamefont {Cho},\ and\ \citenamefont {Oh}}]{Lee}%
  \BibitemOpen
  \bibfield  {author} {\bibinfo {author} {\bibfnamefont {H.~S.}\ \bibnamefont {Lee}}, \bibinfo {author} {\bibfnamefont {H.~B.}\ \bibnamefont {Kim}}, \bibinfo {author} {\bibfnamefont {R.~E.}\ \bibnamefont {Kim}}, \bibinfo {author} {\bibfnamefont {W.~C.}\ \bibnamefont {Ri}}, \bibinfo {author} {\bibfnamefont {B.~K.}\ \bibnamefont {Cho}}, \ and\ \bibinfo {author} {\bibfnamefont {J.~J.}\ \bibnamefont {Oh}},\ }\href@noop {} {\bibfield  {journal} {\bibinfo  {journal} {Journal of the Korean Physical Society}\ }\textbf {\bibinfo {volume} {34}},\ \bibinfo {pages} {88} (\bibinfo {year} {1999})}\BibitemShut {NoStop}%
\bibitem [{\citenamefont {Yahav}\ \emph {et~al.}(2024)\citenamefont {Yahav}, \citenamefont {Potashnikov}, \citenamefont {Pesach}, \citenamefont {Rivin}, \citenamefont {Caspi}, \citenamefont {Rosen}, \citenamefont {Schechter}, \citenamefont {Maniv},\ and\ \citenamefont {Maniv}}]{Preprint_Tb}%
  \BibitemOpen
  \bibfield  {author} {\bibinfo {author} {\bibfnamefont {D.}~\bibnamefont {Yahav}}, \bibinfo {author} {\bibfnamefont {D.}~\bibnamefont {Potashnikov}}, \bibinfo {author} {\bibfnamefont {A.}~\bibnamefont {Pesach}}, \bibinfo {author} {\bibfnamefont {O.}~\bibnamefont {Rivin}}, \bibinfo {author} {\bibfnamefont {E.~N.}\ \bibnamefont {Caspi}}, \bibinfo {author} {\bibfnamefont {J.}~\bibnamefont {Rosen}}, \bibinfo {author} {\bibfnamefont {M.}~\bibnamefont {Schechter}}, \bibinfo {author} {\bibfnamefont {A.}~\bibnamefont {Maniv}}, \ and\ \bibinfo {author} {\bibfnamefont {E.}~\bibnamefont {Maniv}},\ }\href@noop {} {\bibfield  {journal} {\bibinfo  {journal} {(Preprint)}\ } (\bibinfo {year} {2024})}\BibitemShut {NoStop}%
\bibitem [{\citenamefont {Haley}\ \emph {et~al.}(2020)\citenamefont {Haley}, \citenamefont {Weber}, \citenamefont {Cookmeyer}, \citenamefont {Parker}, \citenamefont {Maniv}, \citenamefont {Maksimovic}, \citenamefont {John}, \citenamefont {Doyle}, \citenamefont {Maniv}, \citenamefont {Ramakrishna} \emph {et~al.}}]{EM2}%
  \BibitemOpen
  \bibfield  {author} {\bibinfo {author} {\bibfnamefont {S.~C.}\ \bibnamefont {Haley}}, \bibinfo {author} {\bibfnamefont {S.~F.}\ \bibnamefont {Weber}}, \bibinfo {author} {\bibfnamefont {T.}~\bibnamefont {Cookmeyer}}, \bibinfo {author} {\bibfnamefont {D.~E.}\ \bibnamefont {Parker}}, \bibinfo {author} {\bibfnamefont {E.}~\bibnamefont {Maniv}}, \bibinfo {author} {\bibfnamefont {N.}~\bibnamefont {Maksimovic}}, \bibinfo {author} {\bibfnamefont {C.}~\bibnamefont {John}}, \bibinfo {author} {\bibfnamefont {S.}~\bibnamefont {Doyle}}, \bibinfo {author} {\bibfnamefont {A.}~\bibnamefont {Maniv}}, \bibinfo {author} {\bibfnamefont {S.~K.}\ \bibnamefont {Ramakrishna}},  \emph {et~al.},\ }\href@noop {} {\bibfield  {journal} {\bibinfo  {journal} {Physical Review Research}\ }\textbf {\bibinfo {volume} {2}},\ \bibinfo {pages} {043020} (\bibinfo {year} {2020})}\BibitemShut {NoStop}%
\bibitem [{\citenamefont {Maniv}\ \emph {et~al.}(2021)\citenamefont {Maniv}, \citenamefont {Murphy}, \citenamefont {Haley}, \citenamefont {Doyle}, \citenamefont {John}, \citenamefont {Maniv}, \citenamefont {Ramakrishna}, \citenamefont {Tang}, \citenamefont {Ercius}, \citenamefont {Ramesh} \emph {et~al.}}]{EM}%
  \BibitemOpen
  \bibfield  {author} {\bibinfo {author} {\bibfnamefont {E.}~\bibnamefont {Maniv}}, \bibinfo {author} {\bibfnamefont {R.~A.}\ \bibnamefont {Murphy}}, \bibinfo {author} {\bibfnamefont {S.~C.}\ \bibnamefont {Haley}}, \bibinfo {author} {\bibfnamefont {S.}~\bibnamefont {Doyle}}, \bibinfo {author} {\bibfnamefont {C.}~\bibnamefont {John}}, \bibinfo {author} {\bibfnamefont {A.}~\bibnamefont {Maniv}}, \bibinfo {author} {\bibfnamefont {S.~K.}\ \bibnamefont {Ramakrishna}}, \bibinfo {author} {\bibfnamefont {Y.-L.}\ \bibnamefont {Tang}}, \bibinfo {author} {\bibfnamefont {P.}~\bibnamefont {Ercius}}, \bibinfo {author} {\bibfnamefont {R.}~\bibnamefont {Ramesh}},  \emph {et~al.},\ }\href@noop {} {\bibfield  {journal} {\bibinfo  {journal} {Nature Physics}\ }\textbf {\bibinfo {volume} {17}},\ \bibinfo {pages} {525} (\bibinfo {year} {2021})}\BibitemShut {NoStop}%
\end{thebibliography}
\end{document}